\documentclass[twocolumn,showpacs,amsmath,amssymb,10pt,aps]{revtex4}
\usepackage{graphicx,color}
\usepackage{amsmath}
\usepackage{amssymb}
\usepackage{bbm}
\usepackage{color}

\usepackage[utf8]{inputenc}

\usepackage{amsfonts}
\usepackage{bm}

\newcommand{\ot}{{\,\otimes\,}}
\newcommand{{\Cd}}{{\mathbb{C}^d}}

\newcommand{\bra}[1]{\mbox{$\langle #1 |$}}

\newcommand{\ket}[1]{\mbox{$| #1 \rangle$}}
\def\oper{{\mathchoice{\rm 1\mskip-4mu l}{\rm 1\mskip-4mu l}
{\rm 1\mskip-4.5mu l}{\rm 1\mskip-5mu l}}}
\def\<{\langle}
\def\>{\rangle}

\newtheorem{Cor}{Corollary}

\newtheorem{Remark}{Remark}
\newtheorem{Example}{Example}
\newtheorem{Proposition}{Proposition}

\newcommand{\beq}{\begin{equation}}
\newcommand{\eeq}{\end{equation}}
\newcommand{\bear}{\begin{eqnarray}}
\newcommand{\ear}{\end{eqnarray}}
\newcommand{\bdm}{\begin{displaymath}}
\newcommand{\edm}{\end{displaymath}}

\begin{document}
\title{\textbf{On generalized semi-Markov quantum evolution}}
\author{Dariusz Chru\'sci\'nski and Andrzej Kossakowski}
\affiliation{ Institute of Physics, Faculty of Physics, Astronomy and Informatics \\  Nicolaus Copernicus University,
Grudzi\c{a}dzka 5/7, 87--100 Toru\'n, Poland}



\begin{abstract}
We provide a large class of quantum evolution governed by the memory kernel master equation. This class defines quantum analog of so called semi-Markov classical stochastic evolution. In this Letter for the first time we provide a proper definition of quantum semi-Markov evolution
and using the appropriate gauge freedom  we propose a suitable generalization which contains majority of examples considered so far in the literature.  The key concepts are  quantum counterparts of classical waiting time distribution and survival probability -- quantum waiting time operator and quantum survival operator, respectively. In particular collision models and its generalizations considered recently are special examples of generalized semi-Markov evolution. This  approach allows for an interesting generalization of trajectory description of the quantum dynamics in terms of POVM densities.
\end{abstract}

\maketitle


{\em Introduction.} --- A theory of open quantum systems provides a basic tool to analyze quantum systems which are not isolated  but  interact with  an external environment \cite{Breuer,Weiss,Rivas-Huelga}. Any realistic system is never perfectly isolated and hence this theory  plays a key role for modelling and controlling realistic quantum systems. It is, therefore, clear that open quantum systems are fundamental for potential applications in modern quantum technologies such as quantum communication, cryptography and computation \cite{QIT}.

The standard approach starts with the total ``system + environment" Hamiltonian $H$ and looks for the reduced evolution of the system density operator $\rho_t$ defined by
\begin{equation}\label{RED}
 \rho_0 \ \longrightarrow\ \rho_t = {\rm Tr}_E ( e^{-i H t}\, \rho_0 \ot \rho_E\, e^{iHt}) ,
\end{equation}
where $\rho_E$ is an initial state of the environment and ${\rm Tr}_E$ denotes a partial trace over the environmental degrees of freedom. It is well known that the map $\rho_0 \longrightarrow \rho_t = \Lambda_t[\rho_0]$ is completely positive (CP) and trace-preserving (CPTP) and satisfies $\Lambda_0 = \oper$ (identity map). It is usually called a (quantum) dynamical map. It was shown by Nakajima and Zwanzig \cite{NZ} (see also \cite{NZ-inni,Haake}) that $\rho_t$ satisfies the following generalized master equation
\begin{equation}\label{NZ}
  \dot{\rho}_t = \int_0^t K_{t-\tau} \rho_\tau d\tau ,
\end{equation}
in which quantum memory effects are taken into account
through the introduction of the memory kernel $K_t$. This means that the rate of change of the state $\rho(t)$ at time $t$ depends
on its history (starting at t = 0). The memory kernel is fully determined by the total Hamiltonian and the initial state of the environment. It should be stressed that in general its structure is highly nontrivial since the knowledge of the memory kernel derived from the microscopic model governed by the total Hamiltonian would be in principle equivalent to the knowledge of the full ``system + environment" evolution. Therefore, one may ask about phenomenological memory kernels $K_t$ leading to a legitimate quantum evolution, that is, evolution represented by CPTP map. This is the basic question we address in this Letter: how to characterize physically admissible memory kernels.

Note, that equation (\ref{NZ}) is exact --- it follows from the reduction procedure (\ref{RED}), where one neglects the environmental degrees of freedom (usually the environment lives in the infinite dimensional Hilbert space) and fully cares about the degrees of freedom of the system in question.
To simplify the structure of $K_t$ one may try to apply physically appropriate approximation. Note, however, that approximating $K_t$ is a very delicate issue. A typical  second order Born approximation
considerably simplifies the structure of the memory kernel, however,
in general it violates basic properties of
the master equation -- complete positivity or
even positivity of $\rho_t$.  As is well known Born approximation supplemented by a series of sophisticated Markov approximations results in time-local master equation
 $ \dot{\rho}_t = \mathcal{L}\rho_t$,
with $\mathcal{L}$ being the celebrated GKSL generator \cite{GKS,L}
\begin{equation}\label{Lindblad}
  \mathcal{L}[\rho] = -i[H_{\rm eff},\rho] + \sum_\alpha \gamma_\alpha (V_\alpha \rho V_\alpha^\dagger - \frac 12 \{ V_\alpha^\dagger V_\alpha,\rho\}) ,
\end{equation}
where $H_{\rm eff}$ denotes an effective Hamiltonian, $V_\alpha$ are noise operators, and $\gamma_\alpha \geq 0$ describe decoherence/dissipation rates. On the level of $\mathcal{L}$ one does not care about the microscopic model. Any choice of $H_{\rm eff}$, $V_\alpha$ and $\gamma_\alpha$ leads to legitimate evolution. One would like to find the corresponding characterization on the level of memory kernels. Recently much effort was devoted to non-Markovian quantum evolution which is defined either by time-local generator $\mathcal{L}_t$ or no-local memory kernel $K_t$ (see e.g. recent reviews \cite{Rev1,Rev2}).

The hard problem one faces working with non-local master equation (\ref{NZ}) is how to control complete positivity of the evolution described by the map $\Lambda_t$.   This problem was already faced by Barnett and Stenholm \cite{B-S} for the memory kernel $K_t= k(t)\mathcal{L}$ with $k(t)$ being some memory function  and the legitimate Markovian generator $\mathcal{L}$. An interesting approach of Lidar and Shabani \cite{Lidar} leads to so called {\em post-Markovian master equation} governed by
 $ K_{\rm LS}(t) = k(t) \mathcal{L} e^{ \mathcal{L} t}$.
However, it should be stressed that neither phenomenological kernel of Barnett  Stenholm nor Lindar-Shabani post-Markovian kernel guarnaties  complete positivity of the corresponding dynamical map (see also \cite{Maniscalco,Campbell}). The problem of the admissible memory kernels was then extensively analyzed both from mathematical and physical point of view (see e.g. \cite{Budini,Wilke,AK,B-V,EPL,Wudarski,Bassano-PRL,PRA-2016}). An interesting proposal leading to legitimate memory kernels is provided by so called collision models \cite{C0,C1,C2,C3,C4}. Actually, the non-local memory kernel master eqution is well known for classical stochastic evolution \cite{CLASS,Gil}, where the dynamical map is realized by a family of stochastic matrices.
The aim of this Letter is to provide the quantum analog of classical semi-Markov evolution. Actually, the quantum analog of semi-Markov evolution was already considered by Breuer and Vacchini \cite{B-V}. However, the precise definition was lacking. In this Letter  for the first time we provide a precise definition of quantum semi-markov evolution based on the concepts of quantum waiting time operator and quantum survival operator. Moreover, we show that majority of examples considered so far in the literature fit this class.

\vspace{.2cm}

{\em Quantum evolution from legitimate pairs.} --- In \cite{PRA-2016} we introduced a concept of legitimate pairs, that is, a pair $\{N_t,Q_t\}$ of CP maps such that $N_0 = \oper$ and $Q_t$ satisfies the following constraint: its Laplace transform $\widetilde{Q}_s$ satisfies $||\widetilde{Q}_s||_1 < 1$. Moreover, the following normalization condition has to be satisfied
\begin{equation}\label{nor}
  {\rm Tr}(Q_t[\rho] + \dot{N}_t[\rho])=0 ,
\end{equation}
for all $\rho$. Under these conditions one constructs the dynamical map via
\begin{equation}\label{s}
  \widetilde{\Lambda}_s = \widetilde{N}_s \frac{1}{\oper - \widetilde{Q}_s} = \widetilde{N}_s \sum_{n=0}^\infty \widetilde{Q}_s^{n} \ ,
\end{equation}
or in the time domain
\begin{equation}\label{t}
  \Lambda_t = N_t + N_t \ast Q_t + N_t \ast Q_t \ast Q_t + \ldots ,
\end{equation}
where $A_t \ast B_t = \int_0^t A_\tau B_{t-\tau} d\tau$ denotes the operator convolution. The condition $||\widetilde{Q}_s||_1 < 1$ guaranties that the series in (\ref{s}) converges in the trace norm $||\ ||_1$. By construction $\Lambda_t$ is CP and normalization condition (\ref{nor}) guaranties that $\Lambda_t$ is trace-preserving. The corresponding memory kernel is given by
\begin{equation}\label{L}
  \widetilde{K}_s = (\widetilde{Q}_s -  \oper) \widetilde{N}_s^{-1} + s \oper .
\end{equation}

\begin{Remark}
Actually, due to the fact that in general maps $N_t$ and $Q_t$ do not commute, one may consider another construction
\begin{equation}\label{s1}
  \widetilde{\Lambda}_s =  \frac{1}{\oper - \widetilde{Q}_s} \widetilde{N}_s = \sum_{n=0}^\infty \widetilde{Q}_t^{n}\, \widetilde{N}_s \ ,
\end{equation}
or in the time domain $ \Lambda_t = N_t + Q_t\ast N_t  + Q_t \ast Q_t \ast N_t + \ldots$. The corresponding kernel reads
\begin{equation}\label{R}
  \widetilde{K}_s = \widetilde{N}_s^{-1}(\widetilde{Q}_s -  \oper)  + s \oper .
\end{equation}
In a recent paper \cite{Bassano-PRL} Vacchini calls (\ref{L}) and (\ref{R}) left and right kernels, respectively. In this Letter we follow convention (\ref{s})--(\ref{R}).
\end{Remark}
The simplest example of such pair is provided by $N_t = g(t) \oper$ and $Q_t = f(t) \mathcal{E}$, where $\mathcal{E}$ is a quantum channel, and $f(t),g(t)$ are waiting time distribution and survival probability, respectively,  related by $g(t) = 1 - \int_0^t f(\tau) d\tau$. In this case one finds
\begin{equation}\label{K1}
  \widetilde{K}_s = \frac{\widetilde{f}(s)}{\widetilde{g}(s)}\, (\mathcal{E} - \oper) .
\end{equation}
In this simple example left and right kernels coincide since $N_t$ and $Q_t$ commute.


\vspace{.2cm}

{\em Semi-Markov evolution.} --- Let us recall the construction of the classical semi-Markov evolution \cite{Montroll,Gil,B-V,Esposito,CLASS}: one defines a semi-Markov matrix $q_{ij}(\tau) \geq 0$ for $\tau \geq 0$ such that $\int_0^t q_{ij}(\tau) d\tau$ denotes the probability of jump from state ``$j$" to state ``$i$" no later than  $\tau=t$ provided that at time $\tau=0$ the system stays at the state ``$j$". Now, one defines waiting time distribution $f_j(\tau) =  \sum_i q_{ij}(\tau)$ and survival probability
\begin{equation}\label{}
  g_j(t) = 1 -  \int_0^t f_j(\tau)d\tau ,
\end{equation}
that is the probability that the system stays in the  state ``$j$"  up to $\tau=t$. Clearly $\sum_i \int_0^\infty q_{ij}(\tau) d\tau \leq 1$ and hence $g_j(t) \in [0,1]$ \cite{transient}.
Defining the matrix
\begin{equation}\label{n}
n_{ij}(\tau) = g_j(\tau) \delta_{ij} ,
\end{equation}
the stochastic evolution of the probability vector $p_j(t)$ is realized via the stochastic matrix $T_{ij}(t)$ defined by
\begin{equation}\label{}
  \widetilde{T}(s) = \widetilde{n}(s) \frac{1}{\mathbb{I} - \widetilde{q}(s)} = \widetilde{n}(s)\, \sum_{n=0}^\infty \widetilde{q}^{n}(s) ,
\end{equation}
or in the time domain $T_{ij}(t) = n_{ij}(t) + (n \ast q)_{ij}(t) + (n\ast q \ast q)_{ij}(t) + \ldots$. Moreover, $p_j(t)$
satisfies classical  memory kernel master equation
\begin{equation}\label{C-ME}
  \dot{p}_i(t) = \int_0^t \sum_j [ w_{ij}(\tau) p_j(t-\tau) - w_{ji}(\tau) p_i(t-\tau) ] d\tau  ,
\end{equation}
where the matrix $w_{ij}(t)$ is defined in terms of the Laplace transform as follows
\begin{equation}\label{w-q}
  \widetilde{w}_{ij}(s) = \frac{\widetilde{q}_{ij}(s)}{\widetilde{g}_j(s)} .
\end{equation}
The crucial property of the classical pair of matrices $\{n_{ij}(t),q_{ij}(t)\}$ is that $n_{ij}(t)$ is diagonal and it is uniquely determined by the semi-Markov matrix $q_{ij}(t)$.

Now let us consider quantum case. Let $Q_t$ $(t\geq 0)$ be a family o completely positive maps such that $\int_0^t Q_\tau^\dagger[\mathbb{I}] d\tau \leq \mathbb{I}$ \cite{DUAL}. We call it {\em quantum semi-Markov map} -- a quantum analog of semi-Markov matrix $q_{ij}(t)$. Now, let us define a {\em quantum waiting time operator} $\mathbf{f}_t = Q^\dagger_t[\mathbb{I}]$ and {\em quantum survival operator}
\begin{equation}\label{}
  {\bf g}_t = \mathbb{I} - \int_0^t \mathbf{f}_\tau d\tau .
\end{equation}
It is clear that $ {\bf g}_t  \geq 0$ and $ {\bf g}_0 = \mathbb{I}$. To provide a non-commutative analog of (\ref{n}) let us define a family of CP maps $N_t$ by
\begin{equation}\label{Nt}
  N_t[\rho] = \sqrt{\mathbf{g}_t}\, \rho\,  \sqrt{\mathbf{g}_t} .
\end{equation}
It is clear that $N_0 = \oper$ due to $\mathbf{g}_0 = \mathbb{I}$. Complete positivity of $N_t$ is evident from the Kraus representation (\ref{Nt}). Now, defining the quantum analog of (\ref{w-q})
\begin{equation}\label{}
  \widetilde{W}_s = \widetilde{Q}_s \widetilde{N}_s^{-1} ,
\end{equation}
one finds in the time domain $W_t = Q_t \ast N_t^{-1}$, that is, 
\begin{equation}\label{Wt}
  W_t[\rho] = \int_0^t Q_{t-\tau}[\mathbf{g}_\tau^{-1/2} \rho\, \mathbf{g}_\tau^{-1/2}] d\tau .
\end{equation}
The quantum analog of classical master equation (\ref{C-ME}) reads
\begin{equation}\label{}
  \dot{\rho}_t = \int_0^t K_{t-\tau} \rho_{\tau}\, d\tau ,
\end{equation}
where the memory kernel $K_t$ is constructed as follows: $K_t = W_t - Z_t$ with
\begin{equation}\label{}
  Z_t[\rho] = \frac 12 (W^\dagger_t[\mathbb{I}] \rho + \rho W^\dagger_t[\mathbb{I}]) .
\end{equation}
Again, in the semi-Markov pair $\{N_t,Q_t\}$ the map $N_t$ is fully determined by the map $Q_t$. Hence, the quantum semi-Markov evolution is determined by the semi-Markov map $Q_t$. The characteristic feature of the semi-Markov pair $\{N_t,Q_t\}$ is complete positivity of $N_t^{-1}$. This implies that the map $W_t$ defined in (\ref{Wt}) is CP as well.  Note that the space of semi-Markov maps $Q_t$ is convex, that is, if $Q^{(1)}_t, \ldots, Q^{(k)}_t$ are semi-Markov maps, then for any probability distribution $\{p_i\}$ the map $p_1 Q^{(1)}_t + \ldots + p_k Q^{(k)}_t$ is again semi-Markov.

\vspace{.2cm}

{\em Classical--quantum.} --- Note that a classical semi-Markov evolution immediately follows from the quantum construction if one restricts to the commutative case, that is, one considers
quantum semi-Markov map of the following form
\begin{equation}\label{}
  Q_t[\rho] = \sum_{i,j} q_{ij}(t) |i\>\<j|\rho|j\>\<i| ,
\end{equation}
where $q_{ij}(t)$ is the (classical) semi-Markov matrix. Then the condition $\int_0^t Q_\tau^\dagger[\mathbb{I}] d\tau \leq \mathbb{I}$ implies
$$  \sum_{i}\int_0^t q_{ij}(t) |j\>\<j| \leq \mathbb{I} , $$
which is equivalent to the classical constraint $ \sum_{i}\int_0^t q_{ij}(t) \leq 1$. Moreover, one finds
$$   \mathbf{f}_t = Q^\dagger_t[\mathbb{I}] = \sum_j f_j(t) |j\>\<j| , $$
where $f_j(t)$ stands for a classical waiting time distribution, and
$  \mathbf{g}_t = \sum_j g_j(t) |j\>\<j|$,
with $g_j(t)$ being a classical survival probability. Finally,
$$   {W}_t[\rho] = \sum_{i,j} {w}_{ij}(t) |i\>\<j|\rho|j\>\<i| , $$
with ${w}_{ij}(t)$ defined in (\ref{w-q}).

\vspace{.2cm}

{\em From semi-Markov evolution to Markovian semigroup.} --- It is well known that in the classical case the Markovian semigroup correspond to the specific choice of the semi-Markov matrix $q_{ij}(t)$
\begin{equation}\label{}
  q_{ij}(t) = \pi_{ij} f_j(t) ,
\end{equation}
where $\pi_{ij}$ is the matrix of transition probabilities (a stochastic matrix) and the waiting time distributions $f_j(t)$ read $f_j(t) = \gamma_j e^{-\gamma_j  t}$, with $\gamma_j > 0$. Then one finds for the survival probability $g_j(t) = e^{-\gamma_j  t}$ and finally
\begin{equation}\label{}
  w_{ij}(t) = \delta(t)w_{ij} \ , \ w_{ij}:=\pi_{ij} \gamma_j ,
\end{equation}
which leads to the classical Markovian master equation
\begin{equation}\label{C-M}
  \dot{p}_i(t) = \sum_j [ w_{ij}- \delta_{ij}\gamma_j] p_j(t) .
\end{equation}
In the quantum case one requires that the semi-Markov map has the following structure
\begin{equation}\label{}
  Q_t[\rho] =  \Phi[\sqrt{\mathbf{f}_t}\, \rho\,  \sqrt{\mathbf{f}_t}] \ ,
\end{equation}
where $\Phi$ is an arbitrary quantum channel and the waiting time operator $\mathbf{f}_t$ is given by
\begin{equation}\label{}
  \mathbf{f}_t = \Gamma e^{- \Gamma t} \ ,
\end{equation}
with a positive matrix $\Gamma$. This definition is perfectly consistent: one has $Q_t^\dagger[\mathbb{I}] = \sqrt{\mathbf{f}_t} \Phi^\dagger[\mathbb{I}] \sqrt{\mathbf{f}_t}= \sqrt{\mathbf{f}_t}  \mathbb{I} \sqrt{\mathbf{f}_t}=\mathbf{f}_t  $ due to $\Phi^\dagger[\mathbb{I}]=\mathbb{I}$. One finds $\mathbf{g}_t = e^{-\Gamma t}$ and finally $W_t[\rho] = \delta(t) W$, where
\begin{eqnarray}\label{}
  W[\rho] = 
  \Phi[ \sqrt{\Gamma} \, \rho \, \sqrt{\Gamma}] \ .
\end{eqnarray}
Hence, $ W^\dagger[\mathbb{I}] = \sqrt{\Gamma}\, \Phi^\dagger[\mathbb{I}]\, \sqrt{\Gamma} = \Gamma$,
and one arrives at the following Markovian master equation
\begin{equation}\label{Q-M}
  \dot{\rho}_t = \left( W[\rho_t] - \frac 12 \{\Gamma,\rho_t\} \right) ,
\end{equation}
which is a quantum analog of (\ref{C-M}). Note, that in this case the map $N_t$ is also a semigroup $N_t[\rho] = e^{-\frac 12 \Gamma t} \rho  e^{-\frac 12 \Gamma t}$.

\begin{Remark} Note, that the `Hamiltonian part' $-i[H,\rho]$ is missing in (\ref{Q-M}). It is because we generalized classical construction where such term does nor exist. Interestingly, one may generate Hamiltonian part by a suitable gauge transformation.
\end{Remark}

\vspace{.2cm}

{\em Gauge transformations and generalized semi-Markov evolution.} --- It was proved \cite{PRA-2016} that if $\{N_t,Q_t\}$ provides a legitimate pair then the following maps
\begin{equation}\label{gauge}
  N'_t = \mathcal{G}_t N_t\ , \ \ Q'_t = \mathcal{F}_t Q_t ,
\end{equation}
where $\mathcal{G}_t$ is a dynamical map and $\mathcal{F}_t$ a family of quantum channels, provides another legitimate pair. We call (\ref{gauge}) gauge transformation. Note, that if $Q_t$ is quantum semi-Markov map so is $Q'_t$. Moreover, both $\mathbf{f}_t$ and $\mathbf{g}_t$ are gauge-invariant. Indeed, one has
$$  \mathbf{f}'_t = Q'^\dagger_t[\mathbb{I}] = Q_t^\dagger \mathcal{F}^\dagger_t[\mathbb{I}] =   Q_t^\dagger[\mathbb{I}] =  \mathbf{f}_t \ , $$
due to $\mathcal{F}^\dagger_t[\mathbb{I}] = \mathbb{I}$. It immediately implies $\mathbf{g}'_t = \mathbf{g}_t$.

\begin{Cor} If $\{N_t,Q_t\}$ is a semi-Markov pair, then  $\{N_t,\mathcal{F}_t Q_t\}$ is semi-Markov pair as well with the same waiting time operator $\mathbf{f}_t$ and survival operator $\mathbf{g}_t$.
\end{Cor}

\begin{Remark} On the level of the Markovian master equation (\ref{Q-M}) it means that we change $W$ to $W'=\mathcal{F} W$, where $\mathcal{F}$ is an arbitrary quantum channel. One has $W'^\dagger[\mathbb{I}] =  W^\dagger\mathcal{F}^\dagger[\mathbb{I}]=\Gamma$, that is, $\Gamma$ does not change.
\end{Remark}

\begin{Remark} Suppose that $\{N_t,Q_t\}$ is a Markov pair giving rise to (\ref{Q-M}), that is,
$$   N_t[\rho] = e^{-\frac 12 \Gamma t} \rho  e^{-\frac 12 \Gamma t} \ , \ \ Q_t = W N_t .    $$
Let $\mathcal{G}_t$ be a unitary dynamical map $\mathcal{G}_t[\rho] = e^{-i H t} \rho e^{i Ht}$, where $H$ commutes with $\Gamma$. One finds that $\{ \mathcal{G}_tN_t,Q_t\}$ is a Markov pair giving rise to
\begin{equation}\label{Q-M-H}
  \dot{\rho}_t = -i[H,\rho_t]+\left( W[\rho_t] - \frac 12 \{\Gamma,\rho_t\} \right) ,
\end{equation}
that is, one corrects (\ref{Q-M}) by the Hamiltonian part.
\end{Remark}

Suppose now that $\{N_t,Q_t\}$ is a semi-Markov pair. We call the dynamics constructed out of $\{N'_t,Q'_t\}$ a {\em generalized semi-Markov evolution}. Clearly, generalized semi-Markov evolution is realized by a non-trivial gauge $\mathcal{G}_t$.
It is clear that contrary to the original semi-Markov evolution for the generalized case the map $N_t$ is not entirely determined by $Q_t$. Note, however, that $N_t$ is uniquely determined by $Q_t$ up the a gauge transformation $\mathcal{G}_t$. Therefore, one has

\begin{Proposition} A legitimate pair
$\{N_t,Q_t\}$ corresponds to a generalized semi-Markov evolution if and only if
$$   \mathcal{G}^\dagger_t[a] = \frac{1}{\sqrt{\mathbf{g}_t}}\, N^\dagger[a] \frac{1}{\sqrt{\mathbf{g}_t}} , $$
defines a dual of the legitimate dynamical map, that is, $\mathcal{G}_t$ is CPTP and $\mathcal{G}_0 = \oper$.
\end{Proposition}

\vspace{.2cm}

{\em Generalized semi-Markov evolution vs. generalized collision models.} --- A generalized collision model \cite{C1,C2,C3,C4,Bassano-PRL,PRA-2016} is defined by the following pair
\begin{equation}\label{COL}
  N_t = g(t)\mathcal{G}_t \ , \ \ Q_t = f(t)\mathcal{F}_t \ ,
\end{equation}
where $\mathcal{G}_t$ is an arbitrary dynamical map, $\mathcal{F}_t$ is an arbitrary family of quantum channels (CPTP maps), and $f(t)$, $g(t)$ are waiting time distribution and survival probability, respectively, that is, $g(t) = 1 - \int_0^t f(\tau)d\tau$. Note, that $Q_t$ provides a legitimate quantum semi-Markov map and the corresponding quantum waiting time operator reads $\mathbf{f}_t = Q^\dagger_t[\mathbb{I}] = f(t) \mathbb{I}$. Hence $ \mathbf{g}_t = g(t) \mathbb{I}$. It is, therefore, clear that
$$  N^{\rm SM}_t = g(t) \oper\ , \ \ Q^{\rm SM}_t = f(t)\mathcal{F}_t , $$
defines a semi-Markov pair. Gauging $N^{\rm SM}_t $ by $\mathcal{G}_t$ one obtains (\ref{COL}) which shows that generalized collision model is a special case of generalized quantum semi-Markov evolution.

\begin{Example}[Decoherence] Consider the following qubit quantum semi-Markov map $Q_t[\rho] = \sum_{\alpha=0}^3 p_\alpha f_\alpha(t) \sigma_\alpha \rho \sigma_\alpha$, where $\sigma_\alpha$ are Pauli matrices, $p_\alpha$ is a probability distribution, and real functions $f_\alpha(t) \geq 0$ satisfy $\int_0^\infty f_\alpha(t)dt \leq 1$. These conditions guarantee that $\mathbf{f}_t = \sum_{\alpha=0}^3 p_\alpha f_\alpha(t) \mathbb{I}$ is a legitimate quantum waiting time operator giving rise to the following quantum survival operator $\mathbf{g}_t = g(t) \mathbb{I}$, where
$$ g(t) = 1- \sum_{\alpha=0}^3 p_\alpha \int_0^t f_\alpha(\tau) d\tau . $$
It is clear that $N_t = g(t) \oper$ and the corresponding memory kernel $K_t$ reads
\begin{equation}\label{KP}
  \widetilde{K}_s[\rho] = \sum_{i=1}^3 p_i \frac{\widetilde{f}_i(s)}{\widetilde{g}(s)}\, (\sigma_i \rho \sigma_i - \rho) ,
\end{equation}
and hence it provides direct generalization of (\ref{K1}). Again, in this example left and right kernels coincide. This class of kernels generalizes the class considered in \cite{Wudarski}. This example may be immediately generalized for arbitrary dimension $d$ either by replacing Pauli matrices by unitary Weyl matrices, or by Hermitian Gell-Mann matrices ${\lambda}_\alpha$. In the former case one generalizes (\ref{KP}) to
\begin{equation}\label{KW}
  \widetilde{K}_s[\rho] = \sum_{\alpha=1}^{d^2-1} p_\alpha \frac{\widetilde{f}_\alpha(s)}{\widetilde{g}(s)}\, (U_\alpha \rho U^\dagger_\alpha - \rho) ,
\end{equation}
where $U_\alpha$ are Weyl matrices (cf. \cite{SM}). In the case of Gell-Mann matrices one has $Q_t[\rho] = \sum_{\alpha=0}^{d^2-1} p_\alpha f_\alpha(t) {\lambda}_\alpha \rho {\lambda}_\alpha$, and hence $\mathbf{f}_t = \sum_{\alpha=0}^{d^2-1} p_\alpha f_\alpha(t) {\lambda}_\alpha^2$ which means that $\mathbf{g}_t$ is no longer of the form $g(t) \mathbb{I}$. In this case one has nontrivial map $N_t[\rho] = \sqrt{\mathbf{g}_t} \rho \sqrt{\mathbf{g}_t}$. For the  qutrit case ($d=3$) cf. \cite{SM}.

\end{Example}

\vspace{.2cm}

{\em  Generalized trajectory description.} --- Note the formula (\ref{t}) implies the following  relation \cite{SM}
\begin{equation}\label{POVM}
  \sum_{n=0}^\infty \int_0^t dt_n \int_0^{t_n}dt_{n-1}  \ldots \int_0^{t_2} dt_1 \mathbf{P}^n(t;t_n,\ldots,t_1) = \mathbb{I} ,
\end{equation}
where $\mathbf{P}^0(t) = \mathbf{g}_t$ and
 $$ \mathbf{P}^n(t;t_n,\ldots,t_1) = Q^\dagger_{t_1}  Q^\dagger_{t_2-t_1}  \ldots  Q^\dagger_{t_n-t_{n-1}}[\mathbf{g}_{t-t_n}] . $$
It is clear that $ \mathbf{P}^n(t;t_n,\ldots,t_1) \geq 0$ and due to (\ref{POVM}) they may be considered as POVM densities. Now, if $\rho$ is a density operator, then

\begin{equation*}\label{}
  \rho_t = \sum_{n=0}^\infty \int_0^t dt_n   \ldots \int_0^{t_2} dt_1 p^n(t;t_n,\ldots,t_1) \rho^n(t;t_n,\ldots,t_1)
\end{equation*}
where $  p^n(t;t_n,\ldots,t_1) = {\rm Tr}( \rho \mathbf{P}^n(t;t_n,\ldots,t_1))$, and
$$ \rho^n(t;t_n,\ldots,t_1) = \frac{N_{t-t_n} Q_{t_n-t_{n-1}} \ldots Q_{t_2-t_1} Q_{t_1}[\rho_0] }{ p^n(t;t_n,\ldots,t_1) } , $$
denotes the trajectory with $n$ jumps at $\{t_1,\ldots,t_n\}$.
If $N_t = g(t) \mathcal{G}_t$ and $Q_t = f(t) \mathcal{F}_t$, then $p^n(t;t_n,\ldots,t_1)$
reproduces probability densities for jumps derived by Vacchini \cite{Bassano-PRL}.

\vspace{.2cm}

{\em Conclusions.} --- We provided a precise definition of quantum semi-Markov evolution generalizing well known semi-Markov classical stochastic evolution.
As in the classical case quantum semi-Markov evolution is uniquely defined in terms of quantum semi-Markov map $Q_t$ which gives rise to quantum waiting time operator $\mathbf{f}_t$ and quantum survival operator $\mathbf{g}_t$. Moreover,  using a freedom of gauge transformations, we proposed a suitable generalization which contains majority of examples considered so far in the literature fit this class. In particular collision models studied recently turn out to be generate semi-Markov evolutions. Finally, it has been shown that our  approach allows for an interesting generalization of trajectory description of the quantum dynamics in terms of POVM densities.

\section*{Acknowledgements}

This paper was partially supported by the National Science Center project UMO-2015/17/B/ST2/02026.

\section*{Supplementary material}

In this Supplemental Material we provide technical details for Example 1.

Unitary $d \times d$ Weyl matrices are defined as follows
$$
U_{kl}=\sum_{m=0}^{d-1} \omega^{mk}\ket{m}\bra{m+ l},
$$
with $\omega=e^{2\pi i/d}$, and $k,l=0,1,\ldots,d-1$. They satisfy the well-known relations:
$$
U_{kl}U_{rs}=\omega^{ks}U_{k + r,l + s},\quad U_{kl}^\dag=\omega^{kl}U_{-k,-l}.
$$
Clearly, for $d=2$ the Weyl channel simplifies to the Pauli channel. In what follows we use one index notation $(kl) \longrightarrow \alpha = l + kd = 0,1,\ldots,d^2-1$. Clearly $U_0 = \mathbb{I}$. For $d=3$ one finds

$$   U_1=\left[
\begin{array}{ccc}
 0 & 1 & 0 \\
 0 & 0 & 1 \\
 1 & 0 & 0
\end{array}
\right],\ \   U_2=\left[
\begin{array}{ccc}
 0 & 0 & 1 \\
 1 & 0 & 0 \\
 0 & 1 & 0
\end{array}
\right] , \ \
 U_3=\left[
\begin{array}{ccc}
 1 & 0 & 0 \\
 0 & \omega & 0 \\
 0 & 0 & \omega^2
\end{array}
\right],   $$

$$ U_4=\left[
\begin{array}{ccc}
 0 & 1 & 0 \\
 0 & 0 & \omega \\
 \omega^2 & 0 & 0
\end{array}
\right], \ \
 U_5=\left[
\begin{array}{ccc}
 0 & 0 & 1 \\
 \omega & 0 & 0 \\
 0 & \omega^2 & 0
\end{array}
\right], \ \  U_6=\left[
\begin{array}{ccc}
 1 & 0 & 0 \\
 0 & \omega^2 & 0 \\
 0 & 0 & \omega
\end{array}
\right],$$
$$ U_7=\left[
\begin{array}{ccc}
 0 & 1 & 0 \\
 0 & 0 & \omega^2 \\
 \omega & 0 & 0
\end{array}
\right],\quad U_8=\left[
\begin{array}{ccc}
 0 & 0 & 1 \\
 \omega^2 & 0 & 0 \\
 0 & \omega & 0
\end{array}
\right] \ ,$$
with $\omega = e^{2\pi i/3}$ and $\omega^2 = \omega^* =  e^{-2\pi i/3}$.

Now, let us defined a quantum semi-Markov map

$$  Q_t[\rho] = \sum_{\alpha=0}^{d^2-1} p_\alpha f_\alpha(t) U_\alpha \rho U^\dagger_\alpha , $$
where $p_\alpha$ is a probability distribution, and real functions $f_\alpha(t) \geq 0$ satisfy $\int_0^\infty f_\alpha(t)dt \leq 1$. These conditions guarantee that $\mathbf{f}_t = \sum_{\alpha=0}^{d^2-1} p_\alpha f_\alpha(t) \mathbb{I}$ is a legitimate quantum waiting time operator giving rise to the following quantum survival operator $\mathbf{g}_t = g(t) \mathbb{I}$, where
$$ g(t) = 1- \sum_{\alpha=0}^{d^2-1} p_\alpha \int_0^t f_\alpha(\tau) d\tau . $$
It is clear that $N_t = g(t) \oper$ and the corresponding memory kernel $K_t$ is given by (\ref{KW}).

\vspace{.2cm}

The Gell-Mann matrices for $d=3$ read:

$$ \lambda_1	=	\left[ \begin{array}{ccc} 0& 1& 0 \\ 1& 0& 0\\ 0& 0& 0 \end{array} \right] 	\ , \
\lambda_2	=	\left[ \begin{array}{ccc} 0& -i& 0\\ i& 0& 0\\  0& 0& 0 \end{array} \right] 	\ , \
\lambda_3	=	\left[ \begin{array}{ccc} 1& 0& 0\\ 0& -1& 0\\ 0& 0& 0 \end{array} \right]	 $$

$$ \lambda_4	=	\left[ \begin{array}{ccc} 0& 0 &1 \\ 0& 0& 0\\ 1& 0& 0 \end{array} \right] 	 \ , \
\lambda_5	=	\left[ \begin{array}{ccc} 0& 0& -i\\ 0& 0& 0\\ 1& 0& 0 \end{array} \right] 	\ , \
\lambda_6	=	\left[ \begin{array}{ccc} 0& 0& 0\\ 0& 0& 1\\ 0& 1& 0 \end{array} \right] 	$$

$$ \lambda_7	=	\left[ \begin{array}{ccc} 0& 0& 0\\ 0& 0& -i\\ 0& i& 0 \end{array} \right] 	 \ , \
\lambda_8	=	\frac{1}{\sqrt{3}} \left[ \begin{array}{ccc} 1& 0& 0\\ 0& 1& 0\\ 0& 0& -2 \end{array} \right] 	 . $$
Taking ${\lambda}_0 = \mathbb{I}$ one defines the following quantum semi-Markov map

$$  Q_t[\rho] = \sum_{\alpha=0}^{d^2-1} p_\alpha f_\alpha(t) \lambda_\alpha \rho \lambda_\alpha , $$
where $p_\alpha$ is a probability distribution, and real functions $f_\alpha(t) \geq 0$ satisfy

$$ \int_0^\infty f_\alpha(t)dt \leq || \lambda_\alpha^2 || , $$
where the operator norm  $||\lambda_\alpha^2 ||$ denotes the maximal eigenvalue  $\lambda_\alpha^2$. Note, that all matrices $\lambda_\alpha^2$ are diagonal. These conditions guarantee that $\mathbf{f}_t = \sum_{\alpha=0}^{d^2-1} p_\alpha f_\alpha(t) \lambda_\alpha^2$ is a legitimate quantum waiting time operator giving rise to the following quantum survival operator

$$ \mathbf{g}_t = \mathbb{I} - \sum_{\alpha=0}^{d^2-1} p_\alpha \int_0^t f_\alpha(\tau) \lambda_\alpha^2 d\tau . $$
Hence that map $N_t$ reads $   N_t[\rho] = \sqrt{\mathbf{g}_t} \rho   \sqrt{\mathbf{g}_t}$.
In this case the legitimate pair $\{N_t,Q_t\}$ generates quantum semi-Markov evolution which goes beyond the collision model description.


\begin{thebibliography}{1}
\bibitem{DUAL} A dual map $\Phi^\dagger$ is defined by ${\rm Tr}(A \Phi[B]) = {\rm Tr}(\Phi^\dagger[A] B)$ for all $A, B \in \mathcal{B}(\mathcal{H})$. $\Phi$ is trace-preserving iff its dual is unital, i.e. $\Phi^\dagger[\mathbb{I}] = \mathbb{I}$
\end{thebibliography}
\end{document}